# Potential function of the Wilson – Racah Quantum System

T. J. Taiwo

*Department of Mathematics, University of Benin, Benin City, Edo State 300283, Nigeria*

**Abstract**: In order to establish a correspondence between the reformulation of quantum mechanics without potential function and the convention quantum mechanics, we obtained the potential function of the **New Wilson - Racah quantum system** in [3] using any of the proposed formula in [4]. To achieve this, we used the matrix elements of the potential function and the basis element of the configuration space.

## 1. Introduction

In paper [1] and [2], a new formulation of quantum mechanics without potential function was introduced. Basically, the essence of this reformulation was to enlarge the class of quantum systems. It was observed that there are quantum systems that could be described analytically but their potential function are either difficult to specify or impossible to realize analytically. For instance, the potential functions might be non-analytic, Nonlocal, energy dependent, or the corresponding differential wave equation is higher second order and so on. As a result of this formulation, new quantum systems that do not belong to the conventional solvable class in quantum mechanics were obtained [1-3]

However; in [3], we obtained a new quantum system- **Wilson- Racah Quantum System**withits discrete energy spectrum, scattering phase shift, and discrete wavefunction (Racah quantum system). But for this physical system, we are yet to get the potential function (either analytically or numerically).

As an inverse quantum mechanical problemand with the aim of reconstructing the associated potential function of a physical system; recently in [4] it was confirmed that once thepotential matrix element of the potential function isgotten and with the basis element; the potential function can be derived using any of the four formulas in [4]. This approach (recovering the potential functions) is applicable to all solvable quantum systems in conventional quantum mechanics and the new various quantum systems derived in this new formulation. So in this paper, we seek to get the potential function of the new **Wilson - Racah quantum system**

In section 2; we show briefly how the potential matrix element of the potential function can be derived, and in Section 3 we give a review of the first two methods of how the potential functions can be gotten as shown in [4].

Finally; we plot of the potential function of the Wilson Racah quantum system using its potential matrix elements and the basis element.

## 2. Potential Matrix Elements

From conventional quantum mechanics, the Hamiltonian, $H = T + V$, is sum of the kinetic energy operator $T$ and the potential function $V$. Therefore, we have $V = H - T$. Now to get the potential matrix elements in the basis $\{\varphi_n(x)\}$, we need first to get the matrix elements of the Hamiltonian operator $H$ and kinetic energy operator both in this basis. For one dimension coordinate $x$, $T = -\frac{1}{2}\frac{d^2}{dx^2}$, and also given as $T = -\frac{1}{2}\frac{d^2}{dr^2} + \frac{\ell(\ell+1)}{2r^2}$, in three dimensions with spherical symmetry and radial coordinate $r$ ($\ell$ is the angular momentum quantum number), its matrix representation could be easily derived by operating $T$, the kinetic operator, on the basis elements $\{\varphi_n(x)\}$. In conventional quantum mechanics, the total wave function of the system is $\Psi(t,x) = e^{-iEt/\hbar}\psi(E,x)$; so we have $H\Psi = i\hbar\frac{\partial}{\partial t}\Psi = E\Psi$. Hence, we can therefore write the wave operation as

$$H|\psi\rangle = E|\psi\rangle \tag{1}$$

$H$ is the Hamiltonian operator and $E$ is the energy of the quantum system. However, in the new formulation, the wave function $\psi(E,x)$ is written as

$$\psi_E^\mu(x) = \sum_n f_n^\mu(E)\varphi_n(x) \tag{2}$$

where $\{\mu\}$ is a set of parameters associated with the physical system, $\{f_n^\mu(E)\}$ is parameterized function of the energy, and $\{\varphi_n^\mu(x)\}$ is set of basis set. Writing $f_n^\mu(E) = f_0^\mu(E)P_n^\mu(\varepsilon)$; by the completeness of the basis and energy normalization of the density of state make $\{P_n^\mu(\varepsilon)\}$ a complete set of orthogonal polynomials with positive weight function $\rho^\mu(\varepsilon) = [f_0^\mu(E)]^2$. Hence we have

$$\int \rho^\mu(\varepsilon)P_n^\mu(\varepsilon)P_m^\mu(\varepsilon)d\zeta(\varepsilon) = \delta_{nm} \tag{3}$$

where $\varepsilon$ is proper function of $E$ and $\{\mu\}$; and $d\zeta(\varepsilon)$ is an appropriate energy integration measure. Therefore (2) becomes

$$\psi_E^\mu(x) = \sqrt{\rho^\mu(\varepsilon)}\sum_n P_n^\mu(\varepsilon)\varphi_n(x) \tag{4}$$

Also, we expect all relevant orthogonal polynomials to have asymptotics $(n\to\infty)$ behaviour

$$P_n^\mu(\varepsilon) \approx n^{-\tau}A^\mu(\varepsilon)\cos[n^\xi\theta(\varepsilon) + \delta^\mu(\varepsilon)] \tag{5}$$

where $\tau$ and $\xi$ are real positive constants. $A^\mu(\varepsilon)$ is the scattering amplitude and $\delta^\mu(\varepsilon)$ is the phase shift. Bound states occur as a set (finite or infinite) of zeros of the energies that make the scattering amplitude vanish. So, for a $k^{th}$ bound state at an energy $E_k = E(\varepsilon_k)$ such that $A^\mu(\varepsilon_k) = 0$; the bound state wavefunction becomes

$$\psi_k^\mu(x) = \sqrt{\omega^\mu(\varepsilon_k)}\sum_n Q_n^\mu(\varepsilon_k)\varphi_n(x) \tag{6}$$

$\{Q_n^\mu(\varepsilon_k)\}$ are discrete version of the polynomials $\{P_n^\mu(\varepsilon)\}$ and $\omega^\mu(\varepsilon_k)$ is the associated discrete weight function that satisfy $\sum_k \omega^\mu(\varepsilon_k)Q_n^\mu(\varepsilon_k)Q_m^\mu(\varepsilon_k) = \delta_{n,m}$. Also we can have the wavefunction written as

$$\psi_k^\mu(x,E) = \sqrt{\rho^\mu(\varepsilon)}\sum_n P_n^\mu(\varepsilon)\varphi_n(x) + \sqrt{\omega^\mu(\varepsilon_k)}\sum_n P_n^\mu(\varepsilon_k)\varphi_n(x) \tag{7}$$

this corresponds to the case where there coexist continuous as well as discrete energy spectra simultaneously. In this reformulation, the set of the orthogonal polynomials plays the role of potential function in conventional quantum mechanics and also carry the kinematic information- angular momentum. So in the absence of a potential function, the physical properties of the system are deduced from the features (weight function, nature of generating function, distribution and density of the polynomial zeros, recursion relation, asymptotics, and differential or difference equation) of the orthogonal polynomials $\{P_n^\mu(\varepsilon), Q_n^\mu(\varepsilon_m)\}$.

Also the orthogonal polynomials (continuous and discrete) satisfy three term recursion relations, and produce a tridiagonal matrix representation for the wave operator through the basis set $\{\varphi_n^\mu(x)\}$.

Using (2) in (1) and projecting from the left by $\langle\varphi_n|$ gives the equation $\sum_m P_m^\mu \langle\varphi_n|H|\varphi_m\rangle = E\sum_m P_m^\mu \langle\varphi_n|\varphi_m\rangle$. We can rewrite this as a generalized eigenvalue matrix equation

$$\tilde{H}|P_n^\mu(\varepsilon)\rangle = E\Omega|P_n^\mu(\varepsilon)\rangle, \qquad (8)$$

where $\tilde{H}$ is the matrix representation of the Hamiltonian operator in the basis $\{\varphi_n(x)\}$ and $\Omega$ is the overlap matrix of the basis elements, $\Omega_{n,m} = \langle\varphi_n|\varphi_m\rangle$ (i.e., matrix representation of the identity). We expect (6) should yield the recursion relation $\varepsilon P_n^\mu(\varepsilon) = a_n^\mu P_n^\mu(\varepsilon) + b_{n-1}^\mu P_{n-1}^\mu(\varepsilon) + b_n^\mu P_{n+1}^\mu(\varepsilon)$, for $n = 1, 2, ...$, where $P_0^\mu(\varepsilon) = 1$, $P_1^\mu(\varepsilon) = \alpha\varepsilon + \beta$ and $\{a_n^\mu, b_n^\mu\}$ are the recursion coefficients with $b_n^\mu \neq 0$ for all $n$.

Since all energy polynomials satisfied a three term recursive relation which can be rewritten as in matrix form as $\Sigma|P\rangle = \varepsilon|P\rangle$, where $\Sigma_{n,m} = a_n^\mu \delta_{n,m} + b_{n-1}^\mu \delta_{n,m+1} + b_n^\mu \delta_{n,m-1}$, is a tridiagonal symmetric matrix of the three term recursion relation; therefore the wave equation (8) is equivalent to three term recursion relation of the energy polynomials $\{P_n^\mu(\varepsilon)\}$ and implies that the wave operator matrix $J = \tilde{H} - E\Omega$ will be tridiagonal and symmetric. By this, the matrix representation of Hamiltonian operator can be easily gotten from the three term recursion relation of the energy polynomial and the potential matrix elements of the potential function are easily obtained too.

Since, we always require the matrix wave operator $J = V + T - E\Omega$ be tridiagonal; but if $\Omega$ is non-tridiagonal (i.e., the basis elements are neither orthogonal nor tri-thogonal), then the kinetic energy matrix $T$ will have corresponding energy –dependent components to cancel out the non tridiagonal components. After this, if there still exist further non-tridiagonal components then they will be eliminated by the counter components in $V$.

## 3. The Potential Function Methods

Given $\{V_{nm}\}_{n,m=0}^{N-1}$, where $V_{nm} = \langle\varphi_n|V|\varphi_m\rangle$, as an $N \times N$ matrix elements of the potential function $V(x)$ in a given basis set $\{\varphi_n(x)\}$ and $\{\bar{\varphi}_n(x)\}$ as the conjugate of the basis set ( that is, $\langle\bar{\varphi}_n|\varphi_m\rangle = \langle\varphi_n|\bar{\varphi}_m\rangle = \delta_{nm}$ ); the potential function $V(x)$ of a given system can be gotten numerically. In [4], four different methods were given in computing the potential function making using of the potential matrix element and the basis set. Here we briefly show the derivations of the first two methods (quick practical reference) and we will advise interested readers to check reference [4] for the other two methods. However, in section 4, we used the four methods to confirm the accuracy of the plot.

### 3.1 First Method

Using the Dirac notation,

$$\langle x|V|x'\rangle = V(x)\delta(x-x') \qquad (10)$$

where $\delta(x-x') = \langle x|x'\rangle$, $x$ is the configuration space coordinate, $V$ is the Hermitian operator of the real potential energy, and $V(x)$ is the potential function. Also, we demand $\langle x|\bar{\varphi}_n\rangle = \bar{\varphi}_n(x)$ and $\langle x|\varphi_n\rangle = \varphi_n(x)$. By completeness of the basis set $\sum_n |\bar{\varphi}_n\rangle\langle\varphi_n| = \sum_n |\varphi_n\rangle\langle\bar{\varphi}_n| = I$, the left side and right side of equation (10) can be written as

$$\frac{1}{2}\sum_{n,m=0}^{\infty} V_{nm}\left[\bar{\varphi}_n(x)\varphi_m(x') + \bar{\varphi}_m(x)\varphi_n(x')\right] \tag{11}$$

and

$$\frac{1}{2}V(x)\sum_{n=0}^{\infty} V_{nm}\left[\bar{\varphi}_n(x)\varphi_n(x') + \bar{\varphi}_n(x)\varphi_n(x')\right] \tag{12}$$

Approximately equating (11) to (12) when $x = x'$ for some $N$ (Large enough natural number); we have

$$V(x) \cong \frac{\sum_{n,m=0}^{N-1} V_{nm}\bar{\varphi}_n(x)\bar{\varphi}_m(x)}{\sum_{n=0}^{N-1} V_{nm}\varphi_n(x)\bar{\varphi}_n(x)} \tag{13}$$

This approximation is valid for finite values of $N$ (since $\sum_{n=0}^{\infty}\varphi_n(x)\bar{\varphi}_n(x') = \delta(x-x')$ becomes undefined as $N \to \infty$). Also this formula has been given in [2] without derivation.

### 3.2 Second Method

Here we use the completeness of the configuration space $\int |x'\rangle\langle x'| dx' = 1$ and of the basis set $\sum_n |\bar{\varphi}_n\rangle\langle\varphi_n| = \sum_n |\varphi_n\rangle\langle\bar{\varphi}_n| = I$. For instance, we can therefore write $\langle x|V|\varphi_n\rangle$ as

$$\langle x|V|\varphi_n\rangle = \int\langle x|V|x'\rangle\langle x'|\varphi_n\rangle dx' = V(x)\varphi_n(x) \tag{14}$$

or

$$\langle x|V|\varphi_n\rangle = \sum_{m=0}^{\infty}\langle x|\bar{\varphi}_m\rangle\langle\varphi_m|V|\varphi_n\rangle = \sum_{m=0}^{\infty}\bar{\varphi}_m(x)V_{nm} \cong \sum_{m=0}^{N-1}\bar{\varphi}_m(x)V_{nm} \tag{15}$$

equating (14) to (15), we have

$$V(x) \cong \sum_{m=0}^{N-1}\frac{\bar{\varphi}_m(x)}{\varphi_n(x)}V_{nm} \quad n = 0, 1, .., N-1 \tag{16}$$

In using this formula, only one column ($N$ elements) of the potential matrix element is needed since $V_{nm} = V_{mn}$

### 4. Potential Function of Wilson Racah Quantum System

In [3-Appendix B] in finding the asymptotic (given here in Appendix **A** as A8 or A9) of the Wilson polynomial; we obtained the scattering phase shift of this system as

$$\delta(\varepsilon) = \arg\left[\Gamma(2iy)/\Gamma(\mu+iy)\Gamma(v+iy)\Gamma(a+iy)\Gamma(b+iy)\right] \tag{17}$$

where $y = \varepsilon(E)$ such that $y \geq 0$. Also we choose $\varepsilon = k/\lambda$, where $E = \frac{1}{2}k^2$ and $\lambda^{-1}$ is the length scale of the system in atomic units $\hbar = m = 1$, then $y = \varepsilon = \sqrt{2E}/\lambda$. It was noted that if the all parameters of the Wilson polynomial are positive, there will be no bound states (that is continuous scattering states) while if $\mu < 0$ and $\mu + v$, $\mu + a$, $\mu + b$ are positive then there co-exit the continuum scattering states and $N$ bound states, where $N$ is the largest integer less than or equal to $-\mu$. Also from the zero of the scattering amplitude $A(\varepsilon)$ dictates that the bound states occur at energies $\{\varepsilon_m\}_{m=0}^{N}$ such that $iy = -(m+\mu)$ giving the bound states energy spectrum as

$$E_m = -\frac{\lambda^2}{2}(m+\mu)^2 \tag{18}$$

The total wavefunction for the continuous energy $\varepsilon$ and discrete energy $\varepsilon_m$ can be written as

$$\psi_m(E,x) = \sqrt{\rho^\mu(\varepsilon)}\sum_{n=0}^{\infty} W_n^\mu(\varepsilon^2;v;a,b)\varphi_n(x) + \sqrt{\rho_m^N(\varepsilon)}\sum_{n=0}^{N} W_n^\mu(-(m+\mu)^2;v;a,b)\varphi_n(x) \tag{19}$$

where $\rho^\mu(\varepsilon)$ and $\rho_m^N(\varepsilon)$ are the continuous and discrete normalized weight function from the augmented continuous orthogonality relation of the Wilson polynomial (See Equation 7 in [3]). $W_n^\mu(\varepsilon^2;v;a,b)$ is the Wilson orthogonal polynomial. Also the basis function is chosen in one dimensional configuration space with coordinate $-\infty < x < \infty$ given as $\varphi_n(x) = [\pi 2^{n+m} n!]^{-1/2} e^{-\lambda^2 x^2/2} H_n(\lambda x)$, where $H_n(z)$ is the Hermite Polynomial of degree $n$ in $z$. Now we find the potential function that defined the wavefunction (19) numerically.

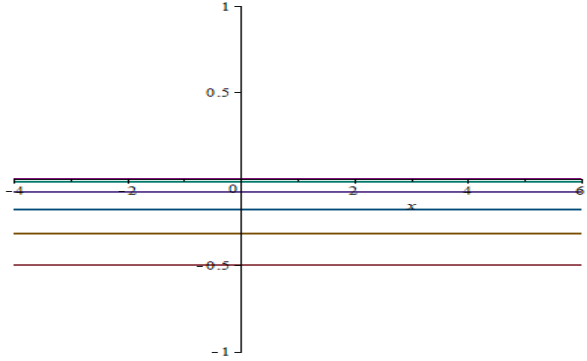

Fig.1. The bound states energy spectrum (8) for the Wilson –Racah quantum system with physical parameters: $\lambda = 0.2$, $\mu = -5$, and $m = 0,1,2,...$

Operating the kinetic operator $T$ on the basis element gives

$$T_{n,m} = -\frac{1}{2}\langle\varphi_n|\frac{d^2}{dx^2}|\varphi_m\rangle = \lambda^2\left(n+\frac{1}{2}\right)\delta_{n,m} - \frac{1}{2}\lambda^2\langle n|(\lambda x)^2|m\rangle \tag{20}$$

where $\langle n|f(z)|m\rangle = [\pi 2^{n+m}n!m!]^{-1/2}\int_{-\infty}^{+\infty} e^{-z^2} f(z)H_n(z)H_m(z)dz$. Since (20) will not be tridiagonal due to the last expression in R.H.S, then we eliminate it by a counter - term in the sought after potential function. So, it is eliminated by the term $+\frac{1}{2}\lambda^4 x^2$, which is a harmonic oscillator potential term. Hence, the potential function we are looking for is

$$V(x) = \frac{1}{2}\lambda^4 x^2 + \tilde{V}(x) \tag{21}$$

The unknown potential function $\tilde{V}(x)$ associated with first term in (10) will be resolved numerically. Using the recursion relation of the Hermite polynomial in (10) without the last term

$$T_{n,m} = -\frac{1}{2}\langle\varphi_n|\frac{d^2}{dx^2}|\varphi_m\rangle = \frac{\lambda^2}{4}\left[(2n+1)\delta_{n,m} - \sqrt{n(n-1)}\delta_{n,m+2} - \sqrt{(n+1)(n+2)}\delta_{n,m-2}\right] \tag{22}$$

which tridiagonal in function space with only odd and or only even indices. The Hamiltonian matrix is gotten by using the following parameters: $y = \sqrt{2E}/\lambda$, $\mu = v$, and $a = b$ in the three term recursion relation of the Wilson orthogonal polynomial (A7)

$$H_{n,m} = \frac{\lambda^2}{4}\left[\left(n+\mu+a-\frac{1}{2}\right)^2 - \left(\mu-\frac{1}{2}\right)^2 - \left(a-\frac{1}{2}\right)^2 + \frac{1}{4}\right]\delta_{n,m}$$

$$-\frac{\lambda^2}{8}\left\{(n+\mu+a-1)\sqrt{\frac{n(n+2\mu-1)(n+2a-1)(n+2\mu+2a-2)}{(n+\mu+a-1)^2-\frac{1}{4}}}\delta_{n,m+1} + (n+\mu+a)\sqrt{\frac{(n+1)(n+2\mu)(n+2a)(n+2\mu+2a-1)}{(n+\mu+a)^2-\frac{1}{4}}}\delta_{n,m-1}\right\}$$

(23)

Hence the matrix elements of the potential function $\tilde{V}(x)$ are $\tilde{V} = H - \tilde{T}$. Finally fig.2 is the full plot of the potential function (21). Using the four methods in [4], we obtained consistency result with the first and third methods. Both gave same results. Below is the graph of the potential function as the potential matrix size $N$ takes value of $N=10$, $N=20$, $N=30$, and $N=50$. Higher order of the potential matrix elements; only shows how the potential function shifts slowly to right hand side. But stability of the potential function is maintained as $N$(order of potential matrix elements) goes from 30 onwards. This potential function encompasses the continuous scattering states and discrete bound states of the Wilson- - Racah quantum system.

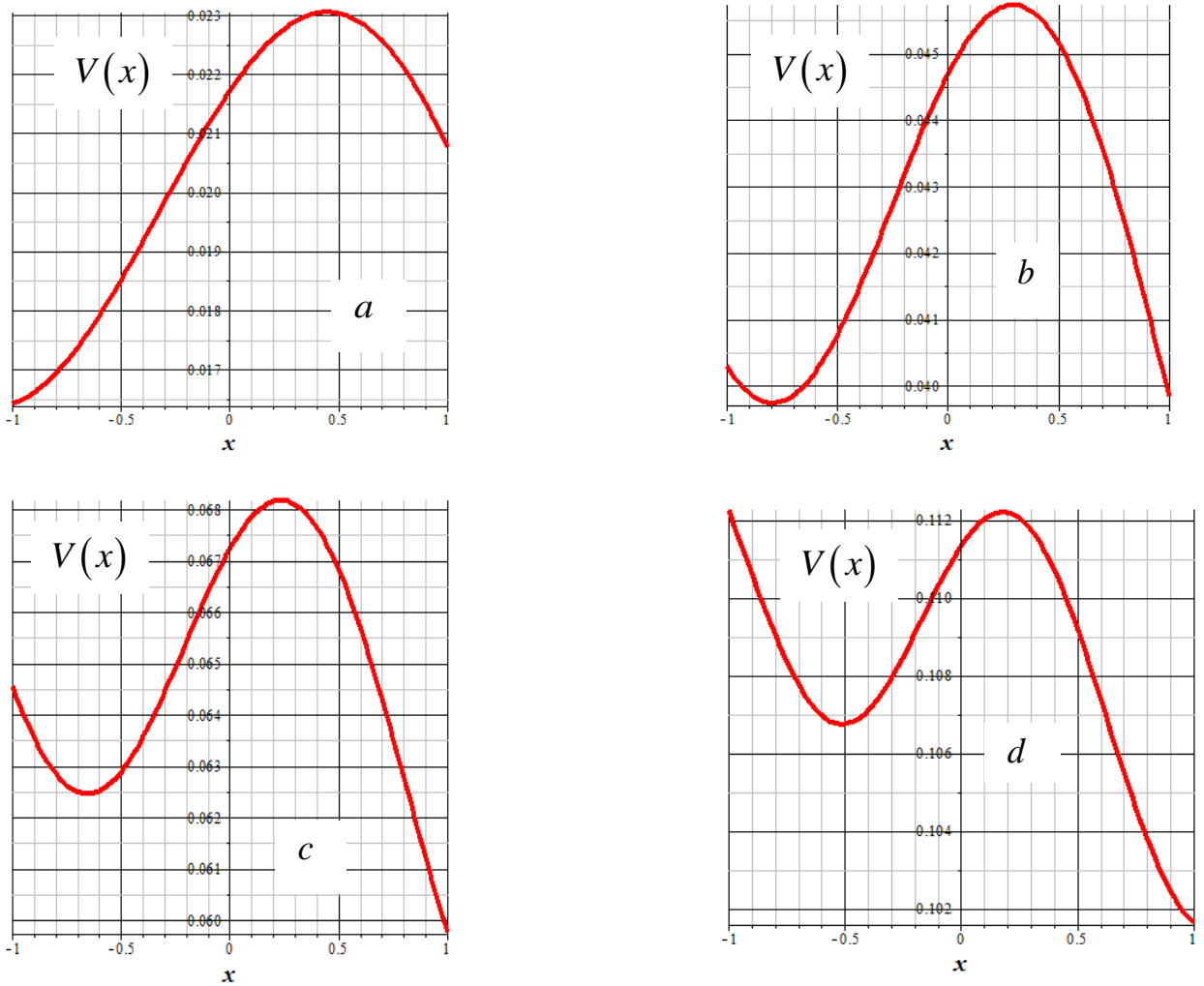

**FIG.2**. The potential function (21) computed with physical parameters: $\lambda = 0.2$ $\mu = v = 0.8$, and $a = 1$..

## 4. Conclusion

In an effort to establish the correspondence between recent reformulation of quantum mechanics without potential function and the convention method; we are able to derive the potential function of a new quantum system –**Wilson -Racah Quantum System**. Also, in this formulation, all solvable quantum systems in the conventional quantum mechanics can be derived through this approach including the new ones, to be best of our knowledge, are unknown in the physics literatures.

## Acknowledgement

The author highly appreciates the support of the Saudi Centre for Theoretical Physics during the progress of this work. Specifically, this paper is dedicated to **Prof. A.D. Alhaidari**.

## Appendix A: The Wilson – Racah Orthogonal Polynomial

Here we give the definitions, properties, asymptotic of the Wilson orthogonal polynomial with its discrete version- Racah polynomial as a review for easy understanding.

The Wilson polynomial, $\tilde{W}_n^\mu(y^2;v;a,b)$ is defined

$$\tilde{W}_n^\mu(y^2;v;a,b) = \frac{(\mu+a)_n(\mu+b)_n}{(a+b)_n n!} \, {}_4F_3\left(\begin{array}{c}-n, n+\mu+v+a+b-1, \mu+iy, \mu-iy \\ \mu+v, \mu+a, \mu+b\end{array}\bigg| 1\right) \tag{A1}$$

where ${}_4F_3\left(\begin{array}{c}a,b,c,d \\ e,f,g\end{array}\bigg| z\right) = \sum_{n=0}^{\infty} \frac{(a)_n(b)_n(c)_n(d)_n}{(e)_n(f)_n(g)_n} \frac{z^n}{n!}$ is the Hypergeometric function and $(a)_n = a(a+1)(a+2)...(a+n-1) = \frac{\Gamma(n+a)}{\Gamma(a)}$. The generating function of these polynomials is

$$\sum_{n=0}^{\infty} \tilde{W}_n^\mu(y^2;v;a,b) t^n = {}_2F_1\left(\begin{array}{c}\mu+iy, v+iy \\ \mu+v\end{array}\bigg| t\right) {}_2F_1\left(\begin{array}{c}a-iy, b-iy \\ a+b\end{array}\bigg| t\right) \tag{A2}$$

Their three term recursion relation $(n=1,2,3...,)$ is

$$y^2 \tilde{W}_n^\mu = \left[\frac{(n+\mu+v)(n+\mu+a)(n+\mu+b)(n+\mu+v+a+b-1)}{(2n+\mu+v+a+b)(2n+\mu+v+a+b-1)} + \frac{n(n+v+a-1)(n+v+b-1)(n+a+b-1)}{(2n+\mu+v+a+b-1)(2n+\mu+v+a+b-2)} - \mu^2\right] \tilde{W}_n^\mu$$

$$- \frac{(n+\mu+a-1)(n+\mu+b-1)(n+v+a-1)(n+v+b-1)}{(2n+\mu+v+a+b-1)(2n+\mu+v+a+b-2)} \tilde{W}_{n-1}^\mu - \frac{(n+1)(n+\mu+v)(n+a+b)(n+\mu+v+a+b-1)}{(2n+\mu+v+a+b)(2n+\mu+v+a+b-1)} \tilde{W}_{n+1}^\mu \tag{A3}$$

The initial seeds $(n=0)$ for this recursion are $\tilde{W}_0^\mu = 1$ and $\tilde{W}_1^\mu = \frac{(\mu+a)(\mu+b)}{(a+b)} - \frac{\mu+v+a+b}{(\mu+v)(a+b)}(y^2+\mu^2)$.

The orthogonality relation of the polynomial is

$$\frac{1}{2\pi} \int_0^\infty \frac{\Gamma(\mu+v+a+b)\left|\Gamma(\mu+iy)\Gamma(v+iy)\Gamma(a+iy)\Gamma(b+iy)\right|^2}{\Gamma(\mu+v)\Gamma(a+b)\Gamma(\mu+a)\Gamma(\mu+b)\Gamma(v+a)\Gamma(v+b)\left|\Gamma(2iy)\right|^2} \tilde{W}_n^\mu(y^2;v;a,b) \tilde{W}_m^\mu(y^2;v;a,b) dy$$

$$= \left(\frac{n+\mu+v+a+b-1}{2n+\mu+v+a+b-1}\right) \frac{(\mu+a)_n(\mu+b)_n(v+a)_n(v+b)_n}{(\mu+v)_n(a+b)_n(\mu+v+a+b)_n n!} \delta_{nm} \tag{A4}$$

The normalized weight function is

$$\rho^\mu(y;v;a,b) = \frac{1}{2\pi} \frac{\Gamma(\mu+v+a+b)\left|\Gamma(\mu+iy)\Gamma(v+iy)\Gamma(a+iy)\Gamma(b+iy)/\Gamma(2iy)\right|^2}{\Gamma(\mu+v)\Gamma(a+b)\Gamma(\mu+a)\Gamma(\mu+b)\Gamma(v+a)\Gamma(v+b)} \tag{A5}$$

The orthonormal version of this polynomial is

$$W_n^\mu(y^2;v;a,b) = \sqrt{\left(\frac{2n+\mu+v+a+b-1}{n+\mu+v+a+b-1}\right)\frac{(\mu+v)_n(a+b)_n(\mu+v+a+b)_n n!}{(\mu+a)_n(\mu+b)_n(v+a)_n(v+b)_n}} \tilde{W}_n^\mu(y^2;v;a,b)$$

$$= \sqrt{\left(\frac{2n+\mu+v+a+b-1}{n+\mu+v+a+b-1}\right)\frac{(\mu+v)_n(\mu+a)_n(\mu+b)_n(\mu+v+a+b)_n}{(a+b)_n(v+a)_n(v+b)_n n!}} {}_4F_3\left(\begin{array}{c}-n,n+\mu+v+a+b-1,\mu+iy,\mu-iy\\ \mu+v,\mu+a,\mu+b\end{array}\middle|1\right) \quad (A6)$$

And the three – term recursion relation for the orthonormal version is

$$y^2 W_n^\mu = \left[\frac{(n+\mu+v)(n+\mu+a)(n+\mu+b)(n+\mu+v+a+b-1)}{(2n+\mu+v+a+b)(2n+\mu+v+a+b-1)} + \frac{n(n+v+a-1)(n+v+b-1)(n+a+b-1)}{(2n+\mu+v+a+b-1)(2n+\mu+v+a+b-2)} - \mu^2\right] W_n^\mu$$

$$-\frac{1}{2n+\mu+v+a+b-2}\sqrt{\frac{n(n+\mu+v-1)(n+a+b-1)(n+\mu+a)(n+\mu+b-1)(n+v+a-1)(n+v+b-1)(n+\mu+v+a+b-2)}{(2n+\mu+v+a+b-3)(2n+\mu+v+a+b-1)}} W_{n-1}^\mu$$

$$-\frac{1}{2n+\mu+v+a+b}\sqrt{\frac{(n+1)(n+\mu+v)(n+a+b)(n+\mu+a)(n+\mu+b)(n+v+a)(n+v+b)(n+\mu+v+a+b-1)}{(2n+\mu+v+a+b-1)(2n+\mu+v+a+b-1)}} W_{n+1}^\mu \quad (A7)$$

The asymptotics $(n \to \infty)$ of this Wilson Polynomial [3 –Appendix B],

$$\tilde{W}_n^\mu(y^2;v;a,b) \approx \frac{2}{n}\Gamma(\mu+v)\Gamma(a+b)|A(iy)|\cos\{2y\ln(n)+\arg[A(iy)]\}+O(n^{-1}) \quad (A8)$$

wherethe scattering amplitude, $|A(iy)|$, is $A(z) = \Gamma(2z)/\Gamma(\mu+z)\Gamma(a+z)\Gamma(b+z)$ when $z = iy$. For the orthonormal version, the asymptotics is

$$W_n^\mu(y^2;v;a,b) \approx B(\mu,v,a,b)\sqrt{\frac{2}{n}}\{2|A(iy)|\cos\{2y\ln(n)+\arg[A(iy)]\}+O(n^{-1})\} \quad (A9)$$

where $B(\mu,v,a,b) = \sqrt{\frac{\Gamma(\mu+v)\Gamma(a+b)\Gamma(\mu+a)\Gamma(\mu+b)\Gamma(v+a)\Gamma(v+b)}{\Gamma(\mu+v+a+b)}}$. The existence of bound states dictates that the scattering amplitude $|A(iy)|$ vanishes i.e. $\mu+iy = -m$, as $m = 0,1,2,...,N$. Hence this implies that $\mu < 0$ and $0 < N \leq -\mu$. Effecting this in the Wilson polynomial changesit to become the**discrete Racah Polynomial** defined as

$$\tilde{R}_n^N(m;\alpha,\beta,\gamma) = \frac{(\alpha+1)_n(\gamma+1)_n}{(\alpha+\beta+N+2)_n n!} {}_4F_3\left(\begin{array}{c}-n,-m,n+\alpha+\beta+1,m-\beta+\gamma-N\\ \alpha+1,\gamma+1,-N\end{array}\middle|1\right) \quad (A10)$$

where $\alpha = \mu+a-1(\alpha > -1)$, $\gamma = \mu+b-1(\gamma > -1)$ $\beta = v+b-1(\beta > N-1)$, and $\delta = -(N+\beta+1) = \mu-b$. However using the inverse parameter map $\mu = \frac{1}{2}(\gamma+\delta+1)$, $v = \beta+\frac{1}{2}(\delta-\gamma+1)$, $a = \alpha-\frac{1}{2}(\gamma+\delta-1)$ and $b = \frac{1}{2}(\gamma-\delta+1)$ takes the Racah polynomial back to the Wilson Polynomial. Also, the generating function of the Racah polynomial is

$$\sum_{n=0}^{N}\tilde{R}_n^\mu(m;\alpha;\beta,\gamma)t^n = {}_2F_1\left(\begin{array}{c}-m,-m+\beta-\gamma\\ -N\end{array}\middle|t\right){}_2F_1\left(\begin{array}{c}m+\alpha+1,m+\gamma+1\\ \alpha+\beta+N+2\end{array}\middle|t\right) \quad (A11)$$

and the three - term recursion relation is

$$\frac{1}{4}(N+\beta-\gamma-2m)^2 \tilde{R}_n^N =$$

$$\left[\frac{1}{4}(N+\beta-\gamma)^2 - \frac{(n-N)(n+\alpha+1)(n+\gamma+1)(n+\alpha+\beta+1)}{(2n+\alpha+\beta+1)(2n+\alpha+\beta+2)} - \frac{n(n+\beta)(n+\alpha+\beta-\gamma)(n+N+\alpha+\beta+1)}{(2n+\alpha+\beta)(2n+\alpha+\beta+1)}\right]\tilde{R}_n^N$$

$$+\frac{(n+\alpha)(n+\beta)(n+\gamma)(n+\alpha+\beta-\gamma)}{(2n+\alpha+\beta)(2n+\alpha+\beta+1)}\tilde{R}_{n-1}^N + \frac{(n+1)(n-N)(n+\alpha+\beta+1)(n+N+\alpha+\beta+2)}{(2n+\alpha+\beta+1)(2n+\alpha+\beta+2)}\tilde{R}_{n+1}^N \quad (A12)$$

The discrete orthogonality relation for the Racah polynomial is

$$\sum_{m=0}^{N} \frac{2m+\gamma-\beta-N}{m+\gamma-\beta-N} \frac{(-N)_m (\alpha+1)_m (\gamma+1)_m (\gamma-\beta-N+1)_m}{(-\beta-N)_m (\gamma-\beta+1)_m (\gamma-\alpha-\beta-N)_m m!} \bar{R}_n^N(m;\alpha,\beta,\gamma) \bar{R}_{n'}^N(m;\alpha,\beta,\gamma)$$

$$= \frac{n+\alpha+\beta+1}{2n+\alpha+\beta+1} \frac{(-\alpha-\beta-N-1)_N (\gamma-\beta-N+1)_N}{(-\beta-N)_N (\gamma-\alpha-\beta-N)_N} \frac{(\beta+1)_n (\alpha+\beta-\gamma+1)_n (\alpha+\beta+N+2)_n n!}{(-N)_n (\alpha+1)_n (\gamma+1)_n (\alpha+\beta+2)_n} \delta_{n,n'} \qquad (A13)$$

where $\bar{R}_n^N(m;\alpha,\beta,\gamma) = {}_4F_3\left(\begin{array}{c}-n,-m,n+\alpha+\beta+1,m-\beta+\gamma-N\\ \alpha+1,\gamma+1,-N\end{array}\bigg|1\right)$ and

$$\rho^N(m;\alpha,\beta,\gamma) = \frac{2m+\gamma-\beta-N}{m+\gamma-\beta-N} \frac{(-N)_m (\alpha+1)_m (\gamma+1)_m (\gamma-\beta-N+1)_m}{(-\beta-N)_m (\gamma-\beta+1)_m (\gamma-\alpha-\beta-N)_m m!}$$

$$\times \frac{(-\beta-N)_N (\gamma-\alpha-\beta-N)_N}{(-\alpha-\beta-N-1)_N (\gamma-\beta-N+1)_N}$$

Similarly like the Wilson polynomial, the discrete Racah polynomial has an orthonormal version defined as

$$R_n^N(m;\alpha,\beta,\gamma) = \sqrt{\frac{2n+\alpha+\beta+1}{n+\alpha+\beta+1} \frac{(-N)_n (\alpha+1)_n (\gamma+1)_n (\alpha+\beta+2)_n}{(\beta+1)_n (\alpha+\beta-\gamma+1)_n (\alpha+\beta+N+2)_n n!}}$$

$$\times {}_4F_3\left(\begin{array}{c}-n,-m,n+\alpha+\beta+1,m-\beta+\gamma-N\\ \alpha+1,\gamma+1,-N\end{array}\bigg|1\right) \qquad (A14)$$

with orthogonality

$$\sum_{m=0}^{N} \rho^N(m;\alpha,\beta,\gamma) R_n^N(m;\alpha,\beta,\gamma) R_{n'}^N(m;\alpha,\beta,\gamma) = \delta_{n,n'} \qquad (A15)$$

Note we made use of the following identities in our calculations: $\frac{(a+1)_n}{(a)_n} = \frac{n+a}{a}$, $\frac{(a)_{n+1}}{(a)_n} = n+a$, $(n+a)_n = \frac{\Gamma(2n+a)}{\Gamma(n+a)}$, $\frac{(n+a)_n}{(a+1)_{2n}} = \frac{a/(a)_n}{2n+a}$, and $(a-n)_n = \frac{\Gamma(a)}{\Gamma(a-n)}$.